\documentclass[prl,twocolumn,aps,superscriptaddress,notitlepage,floatfix,10pt]{revtex4-2}
\usepackage{amsmath,mathtools,amsthm,amssymb,pifont}
\usepackage[utf8]{inputenc}
\usepackage{comment}
\usepackage[american]{babel}
\usepackage{graphicx,xcolor,bbold,titlesec}

\usepackage{MnSymbol}

\usepackage[colorlinks,
bookmarksopen,
bookmarksnumbered,
citecolor=teal,
linkcolor=teal,
urlcolor=teal]{hyperref}
\usepackage{braket}
\usepackage{enumitem}
\usepackage{subfigure}
\usepackage{ifthen}

\usepackage{orcidlink}
\usepackage{tikz,ifthen}
\usepackage{bbold}
\usepackage{tikz-network}
\usetikzlibrary{patterns,decorations.pathreplacing,calligraphy}
\usetikzlibrary{shapes,arrows.meta,decorations.pathmorphing}

\newcommand{\titleinfo}{Mpemba Effects in Quantum Complexity}

\renewcommand{\section}[1]{\textbf{\emph{#1}}.---}

\begin{document}
\title{\titleinfo} 

\author{Sreemayee Aditya~\orcidlink{0000-0002-0412-7944}}
\email{asreemay@uni-koeln.de}
\affiliation{Institut für Theoretische Physik, Zülpicherstraße 77a, 50937, Köln, Deutschland}

\author{Alessandro Summer~\orcidlink{0009-0009-0791-7632}}
\affiliation{School of Physics, Trinity College Dublin, Dublin 2, Ireland}
\affiliation{Trinity Quantum Alliance, Unit 16, Trinity Technology and Enterprise Centre, Pearse Street, D02 YN67, Dublin 2, Ireland}


\author{Piotr Sierant~\orcidlink{0000-0001-9219-7274}}
\affiliation{Barcelona Supercomputing Center Plaça Eusebi Güell, 1-3 08034, Barcelona, Spain}

\author{Xhek Turkeshi~\orcidlink{0000-0003-1093-3771}}
\email{xturkesh@uni-koeln.de}
\affiliation{Institut für Theoretische Physik, Zülpicherstraße 77a, 50937, Köln, Deutschland}

\begin{abstract}
The Mpemba effect is the phenomenon whereby systems farther from equilibrium may relax faster. 
In this work, we show that this counterintuitive behavior appears in the very measures that define quantum complexity. 
Using the framework of quantum resource theories, we study the dynamics of coherence, imaginarity, non-Gaussianity, and magic state resources in random circuit models.
Our results reveal that coherence and imaginarity display a quantum Mpemba effect when the system is initialized in resourceful product states, while non-Gaussianity and magic do not. 
Strikingly, all four resources exhibit the so-called Pontus–Mpemba effect: an initial ``preheating" stage accelerates relaxation compared to direct ``cooling" dynamics. 
Taken together, our findings show that Mpemba physics extends beyond thermodynamics and asymmetry, emerging broadly in the resource theories that capture aspects of  quantum complexity.
\end{abstract}

\maketitle

\section{Introduction}
The Mpemba effect~\cite{mpemba1969cool} is a counterintuitive phenomenon whereby a system can reach equilibrium faster the further it is initially prepared from it. 
This phenomenology is ubiquitous~\cite{teza2025speedupsnonequilibriumthermalrelaxation}: first observed in classical systems~\cite{lasanta2017hotter,lu2017nonequilibrium,klich2019mpemba,kumar2020exponentially,bechhoefer2021fresh,kumar2022anomalous}, it was later extended to open quantum systems~\cite{nava2019lindblad,chatterjee2024multiple,chatterjee2023quantum,kochsiek2022accelerating,carollo2021exponentially,ivander2023hyperacceleration,shapira2024inverse,strachan2025nonmarkovian,zhang2025observation,want2024mpemba,moroder2024thermodynamics,westhoff2025fastdirectpreparationgenuine}. 
A notable recent development is the Pontus-Mpemba effect~\cite{nava2025pontus}, which shows that, for a given initial state, a ``heating" stage, driving the system further from equilibrium, can outperform direct relaxation dynamics. 
Variants of the Mpemba effect have also been established in the non-equilibrium dynamics of closed many-body systems~\cite{Rigol_2008,Polkovnikov_2011,D_Alessio_2016}. 
These so-called quantum Mpemba effects (QMEs) were first identified by analyzing the decay of local asymmetry~\cite{amc-23}, the degree to which a subsystem's reduced density matrix departs from a conserved symmetry, during globally symmetric dynamics: 
the stronger the symmetry breaking in the initial state, the faster it restores the symmetry~\cite{amc-23,ares2023lack,Ferro2024,yamashika2024quantum,Chalas2024,Bertini2024,turkeshi2025quantum,liu2024symmetry,lzyzy-24-2,acm-25}. 
Not only was this experimentally verified in trapped ions~\cite{joshi2024observing}, but it was also extended to a framework beyond symmetry, for instance, studying how fast the initial state reaches the stationary one under Hamiltonian dynamics~\cite{ares2025simplerprobequantummpemba,Bhore2025mpemba}. 


\begin{figure}[t!]
\includegraphics[width=\columnwidth]{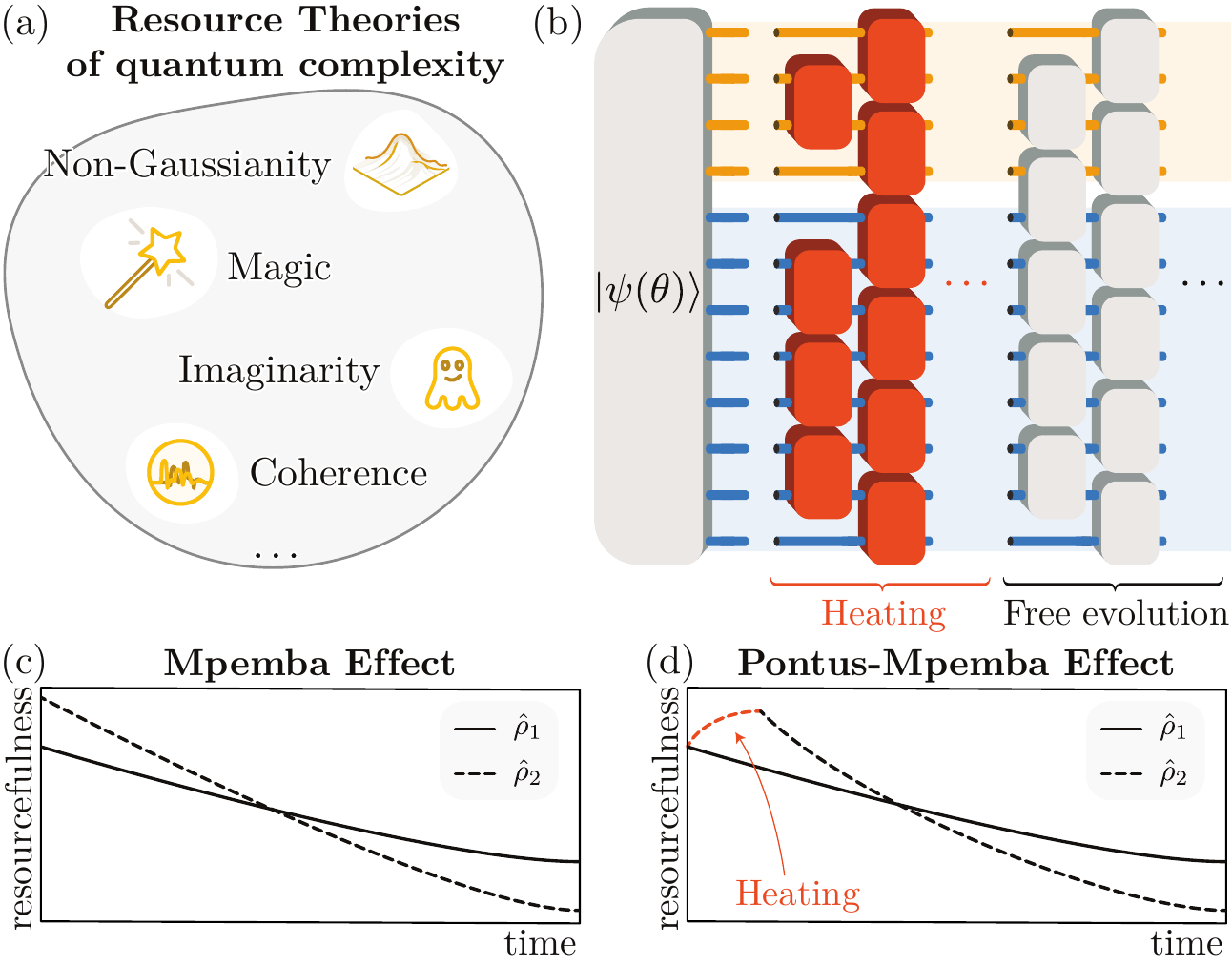}
\caption{ 
(a) \textit{Quantum Resources}: We study coherence, imaginarity, non-Gaussianity and magic resources to extend the Mpemba effect to measures of quantum complexity.
(b) \textit{Setup}: a resourceful state $|\Psi(\theta)\rangle$ is prepared, with $\theta$ controlling its resource content. For the quantum Mpemba effect (QME), different initial states evolve under free operations (grey), and resources dissipate locally in subsystem $A$ (light orange) while remaining conserved globally. In the Pontus-Mpemba effect, the same state is first ``heated'' in $A$ and its complement before free evolution.
(c) \textit{Quantum Mpemba Effects}: A more resourceful state dissipates faster than a less resourceful one.
(d) \textit{Pontus-Mpemba Effects}: Preheating accelerates local dissipation compared to free evolution alone. 
}
\label{fig:schematic}
\end{figure}

Recently, Ref.~\cite{summer2025resourcetheoreticalunificationmpemba} put these results, from classical to quantum setups, within the same umbrella of resource theories~\cite{RT2,deneris2025analyzingfreestatesquantum,diaz2025unifiedapproachquantumresource}. 
In this picture, inspired by thermodynamics, one identifies free states (e.g., thermal or symmetric states) and free operations (dynamics that cannot generate resources), while resourceful states are those that deviate from these constraints. 
A Mpemba effect arises when two such states evolve under the same free dynamics and the more resourceful one dissipates its resource faster, leading to a crossing of resource monotones. For example, the thermal Mpemba effect corresponds to the resource theory of athermality, while the symmetry Mpemba effect is fully captured by asymmetry resources.


Here, we push these ideas in a 
radically different direction by demonstrating that Mpemba physics can arise in the metrics of quantum complexity--ranging from coherence~\cite{coherence3,coherence2,coherence1} and magic resources~\cite{bravyi2005universal,magic2,magic6,leone2022stabilizer} to non-Gaussianity~\cite{lumia2024measurementinduced,hebenstreit2019all,lyu2024displacedfermionicgaussianstates,sierant2025fermionicmagicresourcesquantum} and imaginarity~\cite{imag1,imag2,imag3,imag5}. 
Unlike thermodynamic observables or symmetry-based that were quantities considered in Ref.~\cite{summer2025resourcetheoreticalunificationmpemba}, these resources do not admit a direct physical interpretation, but instead quantify irreducible aspects of quantum complexity. 
Nevertheless, our results show that even such abstract resources can strikingly display Mpemba physics: \textit{more resourceful states may dissipate their local quantum complexity content faster than less resourceful ones under free unitary evolution. }
We substantiate this fact by tracking resource monotones in subsystems of closed quantum systems evolved under brickwork circuits of random two-body gates, starting from a tilted polarized, or ferromagnetic, state.
We find that coherence and imaginarity exhibit a QME in this context, whereas magic and non-Gaussianity do not. 
Strikingly, a resourceful preheating stage induces a quantum Pontus–Mpemba effect (QPME) for all resources, with relaxation accelerated compared to direct resource free evolution from the same initial state. 
Taken together, our results demonstrate that the QME transcends thermodynamic settings and manifests directly in the very quantities that define quantum complexity.

\section{Quantum Resources and Mpemba Effects}
We begin with a brief overview of quantum resource theories~\cite{RT2}, which offer a natural framework to study Mpemba phenomenology through measures of quantum complexity.  
Quantum resource theories provide a unifying language for characterizing and quantifying nonclassical features across diverse physical settings [cf. Fig.~\ref{fig:schematic}(a)]. 
The central idea is to distinguish between free states, which can be prepared without cost, and free operations, which cannot generate resource from these states. Any state outside the free set is deemed resourceful. 

This abstract framework encompasses a wide variety of phenomena~\cite{RT2}. In thermodynamics, the relevant resource is athermality: the free states belong to the Gibbs ensembles, the free operations are thermal maps that cannot generate athermality, and any out-of-equilibrium state is resourceful. In systems with symmetries, the resource is asymmetry, with free states invariant under the symmetry group, free operations given by covariant channels, and resourceful states are those that break the symmetry. In quantum information theory, coherence and entanglement serve as resources: here the free states are respectively basis-diagonal
or separable states, the free operations are incoherent or local operations and classical communication, and any state outside these sets is resourceful.
Beyond these familiar examples, QRTs also capture more abstract forms of quantumness central to quantum complexity. In the resource theory of magic, the free states are stabilizer states, the free operations are the Clifford unitaries~\cite{Haug2023stabilizerentropies, leone2024stabilizer}, and all magic, or non-stabilizer, states--enabling universal quantum computation~\cite{bravyi2005universal}--are resourceful. 
For non-Gaussianity, the free states are Gaussian states of fermionic modes, the free operations are fermionic Gaussian unitaries, 
while interactions between fermions may increase the resourcefulness.

To quantify resourcefulness, one introduces resource monotones $M(\rho)$, functions that satisfy two basic requirements: (i) $M(\rho)=0$ for all free states $\rho$, and (ii) $M(\Lambda[\rho]) \leq M(\rho)$ for all free operations $\Lambda$. 
A simple class of such measures is defined through the minimal distance of a state from the set of free states. 
While conceptually clear, these definitions are typically intractable in many-body settings, as they involve 
optimizations that become prohibitive beyond a few qubits; see, e.g., ~\cite{Hamaguchi2024handbook}.
Throughout this work we therefore focus on scalable monotones, detailed in the following, that provide faithful diagnostics of quantumness in the many-body
systems.

Although the framework is broadly applicable, we focus, for later convenience, on the setups most relevant to this work and define the QME [cf. Fig.~\ref{fig:schematic}(b)]. 
Consider two initial $N$-qubits states $|\Psi_1\rangle$ and $|\Psi_2\rangle$, resourceful with respect to a given quantum resource theory $\mathfrak{Q}$, evolving under the same dynamics $U_t$ composed solely of free operations. 
In this setting, the global resource content is conserved by construction, but the reduced states on a subsystem $A$, $\rho_{1/2}(t) = \mathrm{Tr}_{B}\!\left[U_t |\Psi_{1/2}\rangle\langle \Psi_{1/2}| U_t^\dagger \right]$, follow a dynamics that is no longer free. 
When the subsystem $A$ is much smaller than its complement $B$, it can effectively dissipate its resources into $B$, which acts as a reservoir.
A QME occurs when the density matrices $\rho_{1/2}(t)$, satisfy two conditions: (i) $M(\rho_1(0)) > M(\rho_2(0))$ at the initial time, and (ii) $M(\rho_1(t)) < M(\rho_2(t))$ after some timescale $t>\tau$ [cf. Fig.~\ref{fig:schematic}(c)]. 
In essence, 
the QME occurs when the local resource content can be dissipated faster when the initial state is more resourceful.
Similarly, the 
QPME arises in a complementary setting [cf. Fig.~\ref{fig:schematic}(d)]~\cite{nava2025pontus}: starting from the same state $|\Psi\rangle$, an additional resourceful ``heating'' stage $\tilde{U}_T$, followed by free evolution $U_t$, can lead to a faster dissipation of resources than direct free evolution over the timescale $U_{t+T}$. 
Concretely, this occurs when (i) $M(\tilde{\rho}(T)) > M(\rho(T))$ after the heating stage, yet (ii) $M(\tilde{\rho}(T+\tau)) < M(\rho(T+\tau))$ at later times, where $\rho_t = \mathrm{Tr}_{B}\!\left[U_t |\Psi_{1/2}\rangle\langle \Psi_{1/2}| U_t^\dagger \right]$ and 
\begin{equation}
    \tilde{\rho}_t\equiv \begin{cases}
        \mathrm{Tr}_{B}\!\left[\tilde{U}_t |\Psi_{1/2}\rangle\langle \Psi_{1/2}| \tilde{U}_t^\dagger \right] & t\le T\\ 
        \mathrm{Tr}_{B}\!\left[U_{t-T} \tilde{U}_T |\Psi_{1/2}\rangle\langle \Psi_{1/2}| \tilde{U}_T^\dagger U_{t-T}^\dagger\right] & t> T\;.
    \end{cases}
\end{equation}
In the following sections, we demonstrate how these phenomena arise within the resource theories of coherence, imaginarity, non-Gaussianity, and magic resources. 
Additional results are also presented in the End Matter.

\begin{figure*}[htb]
\centering
\includegraphics[width=0.25\textwidth]{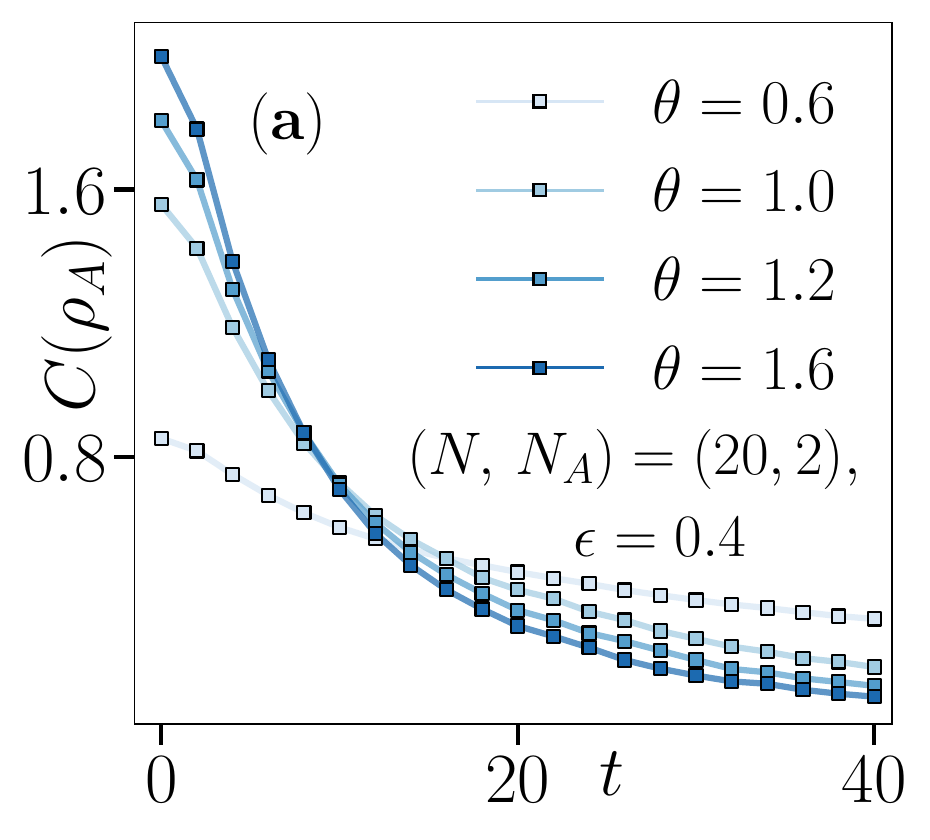}\hspace{-0.005\hsize}%
\includegraphics[width=0.25\textwidth]{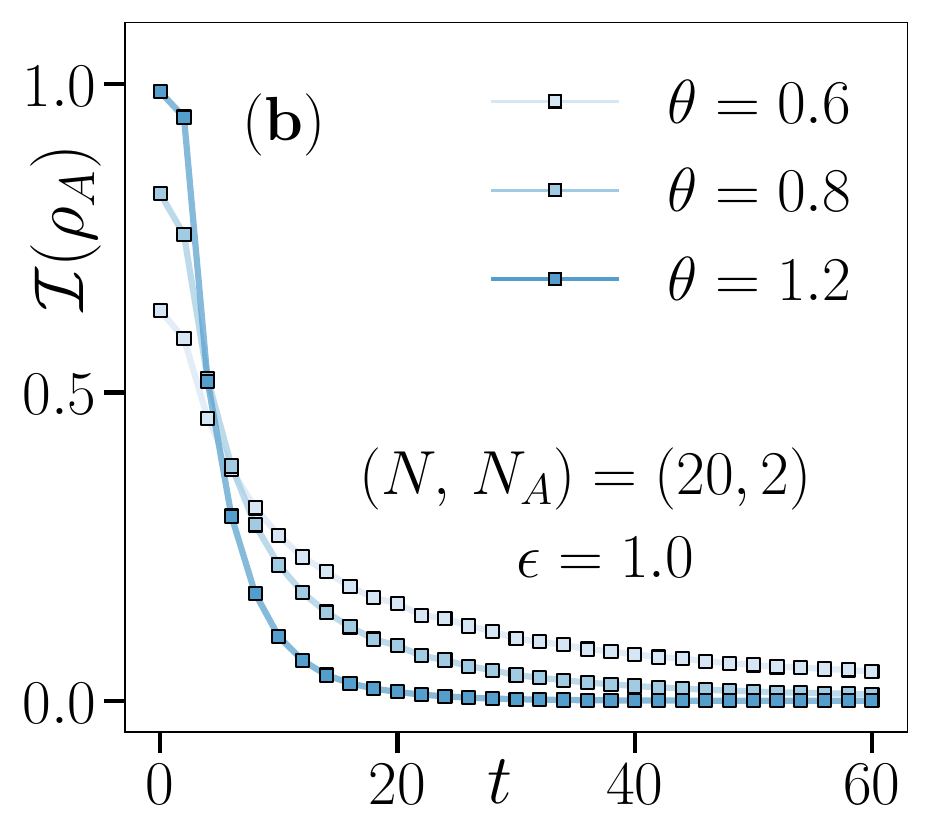}\hspace{-0.005\hsize}%
\includegraphics[width=0.25\textwidth]{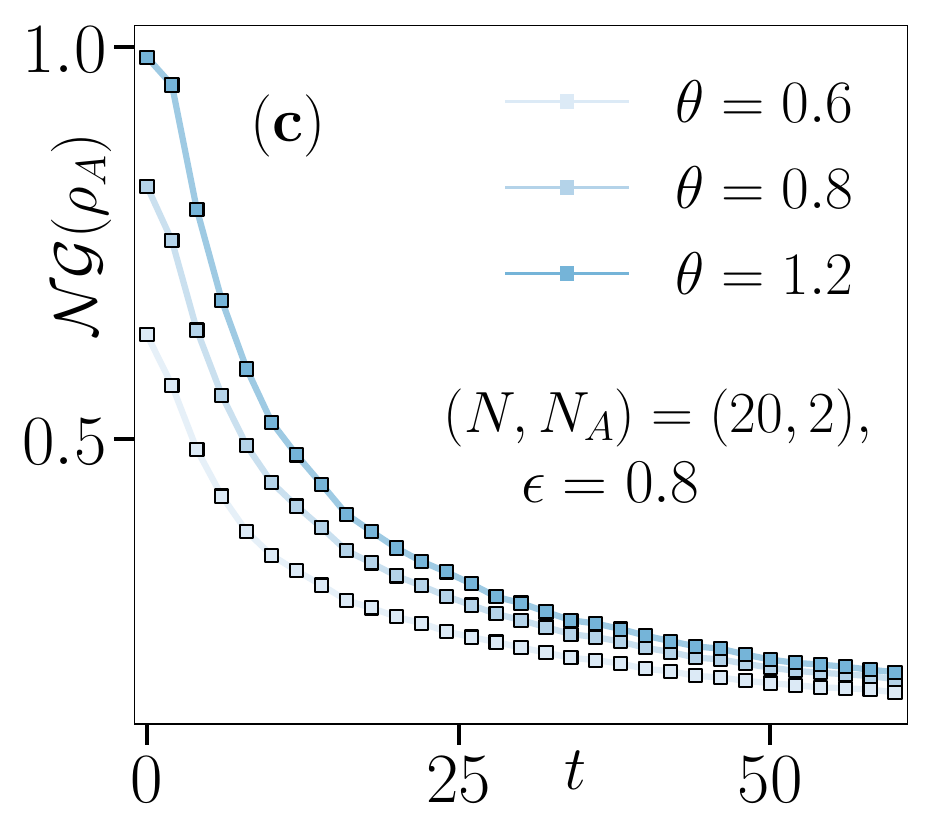}\hspace{-0.005\hsize}%
\includegraphics[width=0.25\hsize]{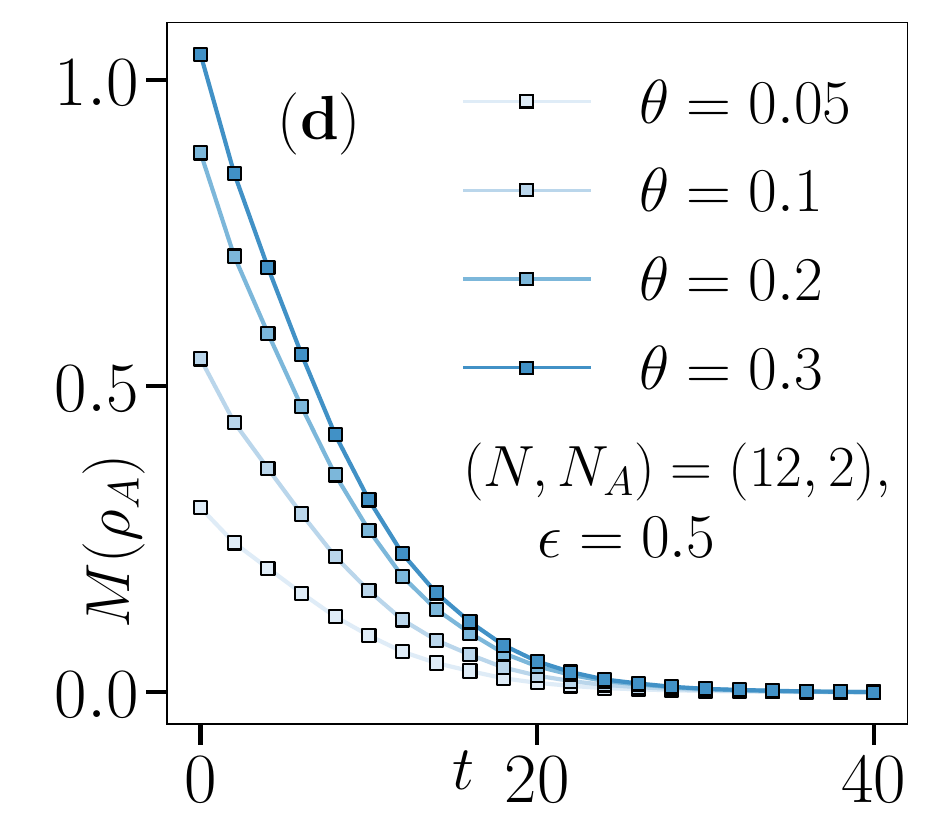}\hspace{0.0\hsize}
\caption{
The QME for quantum resources in
a 1D chain of $N=20$ 
qubits: 
coherence (a), imaginarity (b), and non-Gaussianity (c); and for $N=12$ qutrits for quantum magic resources (d). 
The resource content of a subsystem of size $N_A=2$ is evaluated with the suitable resource monotones, starting from 
the initial state $\ket{\Psi(\theta}$. The dynamics are governed by the
circuit $U_t$ comprising free unitary operations, and 
the results are averaged over $4000$ 
circuit realizations. While these choices of initial states reveal the characteristic crossings of the QME for various tilting angles, $\theta$,
for the coherence and imaginarity, see
Fig. (a-b), no comparable effect is observed in the case of non-Gaussianity and magic resources, as shown in Fig. (c-d).}
\label{fig1:QME}
\end{figure*}

\section{Quantum Mpemba Effects}
To consider different resources on equal footing, we adopt a common circuit-based setup. 
Specifically, we consider a system of $N$ qudits initialized in a resourceful state $|\Psi(\theta)\rangle$, where the continuous parameter $\theta$ controls the degree of resourcefulness of the initial state. 
Throughout this work, we take  $|\Psi_{O}(\theta)\rangle=e^{-i \theta \sum_{i} O_i}|0\rangle^{\otimes N}$ with local onsite operators $O_i$ chosen such that the resulting states are resourceful for the respective theories. 
The dynamics is generated by a random brickwork circuit, $U_t =\prod_{s=1}^t u_s$ with $u_s=\prod_{x\in \mathcal{B}_s} u_{s,x,x+1}$, where $\mathcal{B}_{s}$ denotes the set of odd (even) sites for odd (even) $s$, and each two-qubit gate is drawn independently from a subgroup of free operations $u_{s,x,x+1}\sim \mathcal{G}\subset \mathcal{U}_\mathrm{free}$. 
To reduce the discrete-time effect and control the density of 
unitary operations to obtain a desirable Mpemba time, the gates are applied stochastically: with probability $\epsilon$, $u_{s,x,x+1}$ is drawn from a subgroup of free operations $\mathcal{G}$, while with probability $1-\epsilon$, it acts trivially as $\mathbb{1}_{x,x+1}$. 
Further, in the following, we will consider either periodic boundary condition or open boundary conditions.
We analyze the bipartition defined by the interval $A=\{1,\dots,N_A\}$ and its complement $B$.
Different setups are specified by two ingredients: 
(i) the choice of gate ensemble $\mathcal{G}$, and (ii) the resource monotone $M$.

We begin our analysis with the resource theory of coherence~\cite{coherence3,coherence2,coherence1}. As initial states we consider the class of tilted ferromagnets, $\ket{\Psi_{Y}(\theta)}=e^{-i \theta/2\sum_{i} Y_i}|0\rangle^{\otimes N}$~\footnote{We define $X,Z$ are the Pauli matrices for $d=2$, and the generalized Pauli operators for $d>2$ as $Y = iZX$.}, which are resourceful since they form coherent superpositions of multiple computational basis states. Larger values of the tilt angle $0\le \theta \le \pi/2$ correspond to more resourceful states.  
To quantify the coherent resources of subsystem $A$, we use the relative entropy of coherence,
\begin{equation}
    C(\rho_A) = S(\rho_A^D) - S(\rho_A),
    \label{eq:rel_ent_coh}
\end{equation}
where $S(\rho_A) = -\mathrm{Tr}(\rho\log_2 \rho)$ and $\rho_A^D = \sum_{z} \langle z|\rho_A|z\rangle |z\rangle\langle z|$ is the dephased (diagonal) part of $\rho_A$ in the computational basis.
The circuit is built from phase-dressed permutation gates $u\in\mathcal{G}_C$, of the form $u=\exp[-i \sum_{a,b=0}^1 \alpha_{a,b} Z^{a}\otimes Z^b] S$, with uniform random phases $0\le \alpha_{a,b}\le 2\pi$ and $S$ either the identity or the SWAP gate. 
These gates preserve the coherence of the global state while still redistributing, and thereby altering, the coherence content locally.  
The resulting random circuit dynamics ensures that, in the stationary limit, the reduced density matrix $\rho_A$ becomes diagonal whenever $N_A \ll N$ with $N\gg 1$, i.e., in the thermodynamic limit. 
At late times, the subsystem dissipates its entire resource content, with $C(\rho_A)\to 0$. 
We simulate the dynamics using state-vector methods for $N=20$ and $N_A=2$, fixing $\epsilon=0.4$ and averaging over $\mathcal{N}=4000$ circuit realizations; the results are shown in Fig.~\ref{fig1:QME}(a).
As $\theta$ increases, the initial state becomes more resourceful, yet its coherence decays faster with circuit depth (time). The average relative entropy of coherence exhibits characteristic crossings at intermediate times--a hallmark of QME. 
In the End Matter, we analyze Clifford circuits and demonstrate an analogous QME for coherence in large systems of up to $N=512$ qubits.

Next, we turn our attention to the resource theory of imaginarity~\cite{imag1,imag2,imag3,imag4,imag5}.
This treats quantum states with purely real amplitudes as free, and quantifies the extent to which complex phases provide a nonclassical resource.
We quantify imaginarity using the relative entropy of imaginarity, 
\begin{equation}
\mathcal{I(\rho)}=S(\mathfrak{R}(\rho))-S(\rho),
\label{eq:rel_ent_imag}
\end{equation}
where $\mathfrak{R}(\rho)$ denotes the elementwise real part of $\rho$, i.e., $\langle m|\mathfrak{R}(\rho)|n\rangle = \mathrm{Re}[\langle m|\rho|n\rangle]$.
The circuit evolution is generated by two-body free (orthogonal) gates $u\in\mathcal{G}_I$ of the form $u=\exp[-i \alpha (X\otimes Y - Y\otimes X)/2]$ where $\alpha$ is a random phase uniformly sampled from $[0,2\pi]$.
The numerical results are shown in Fig.~\ref{fig1:QME}(b) for a chain of $N=20$ qubits and a subsystem of size $N_A=2$, averaged over $\mathcal{N}=4000$ realizations of the circuit with $\epsilon=1$, starting from the $x$-tilted initial state $\ket{\Psi_{X}(\theta)}=e^{-i\theta/2\sum_{j}X_{j}}\ket{0}^{\otimes N}$. 
The evolution of the relative entropy of imaginarity exhibits the hallmark crossing behavior: states with larger $\theta$ start more resourceful but dissipate their imaginarity content faster with circuit depth, providing clear evidence of the QME.

Not all resources exhibit the QME starting from these tilted ferromagnetic states.  
We showcase this point with two examples: the resource theory of fermionic non-Gaussianity~\cite{lumia2024measurementinduced, hebenstreit2019all, lyu2024displacedfermionicgaussianstates, sierant2025fermionicmagicresourcesquantum} and the magic one~\cite{veitch2012negative}. 
For the fermionic non-Gaussianity it is convenient to define the $2N$ Majorana operators via the Jordan-Wigner transformation $\gamma_{2m-1}=(\prod_{i=1}^{m-1} Z_i) X_m$ and $\gamma_{2m}=(\prod_{i=1}^{m-1} Z_i) Y_m$, which satisfy the anticommutation relation $\{\gamma_{a},\gamma_b\}=2\delta_{a,b}$. 
Then we quantify the non-Gaussianity via the relative entropy of Gaussianity~\cite{lumia2024measurementinduced}
\begin{equation}
    \mathcal{NG}(\rho_A) = S_{\mathcal{G}}(\rho_A) - S(\rho_A),
    \label{eq.:rel_ent_ng}
\end{equation}
where $S(\rho_A)$ is the von Neumann entropy. The entropy for a Gaussian reference state, $\mathcal{S}_G(\rho_A)$ in Eq. \eqref{eq.:rel_ent_ng} can be computed in terms of the non-negative eigenvalues of the 
$2N_A \times 2N_A$ Majorana correlation matrix $M(\rho_A)$ as
 $\sum_{j=1}^{N_A} H\Big(\frac{1+\lambda_j}{2}\Big)$,
where $H(x) = -x \log_2 x - (1-x) \log_2 (1-x)$ is the binary entropy function, and $\lambda_j \geq 0$ are the non-negative eigenvalues of $M(\rho_A)$, whose elements are defined by 
$M_{a,b}(\rho_A) = -\frac{i}{2}\mathrm{Tr}\big(\rho_A [\gamma_a, \gamma_b]\big)$, 
with the Majorana operators $\gamma_a$.
In this setup, the free evolution is built out of fermionic Gaussian gates $u\in \mathcal{G}_\mathrm{NG}$ specified by a random anti-symmetric matrix $H$ as $u=\exp[-i \sum_{a,b=1}^4 H_{a,b} \gamma_a \gamma_b]$. Our numerical results, shown in Fig.~\ref{fig1:QME}(c) for $N=20$ and subsystem size $N_A=2$, are averaged over $\mathcal{N}=4000$ random circuit realizations with $\epsilon=0.8$. While the steady state is again a free fermionic state in the thermodynamic limit, the dynamics of $\mathcal{NG}$ shows no crossings. 

Lastly, we turn to magic state resources~\cite{veitch2012negative,magic2,magic3,magic4,magic5,magic6} and consider a one-dimensional chain of $N$ qutrits ($d=3$ local dimension). 
Denoting the generalized Pauli matrices $Z=\sum_{m=0}^{d-1} e^{2\pi i m /d}|m\rangle\langle m|$ and $X=\sum_{m=0}^{d-1} |m\rangle\langle m+1\;\mathrm{mod}\;d|$, we define the Pauli strings $P_r = \prod_{l=1}^{N} e^{\pi i (d+1)r_{l}^{x} r_{l}^z/2} X_{l}^{r_{l}^{x}}Z_{l}^{r_{l}^z}$ via $r=(r_{1}^{x},r_{1}^{z},\cdots,r_{N}^{x},r_{N}^{z})\in \mathbb{Z}_{d}^{2N}$.
These allow us to characterize magic resources through the mana~\cite{veitch2012negative,Tarabunga2024criticalbehaviorsof,turkeshi2025magic},
\begin{equation}
M(\rho) = {\log_2}\left(\sum_{r\in \mathbb{Z}_d^{2N}}|\mathcal{W}_{r}|\right),
\label{eq:mana}
\end{equation}
where $A_0= \sum_{r\in \mathbb{Z}_{d}^{2N}} P_r/d^N$ and $A_r = P_r A_0 P_r^\dagger$~\cite{Gross2006Hudson}. 
Starting from the tilted polarized state $|\Psi_{X}(\theta) \rangle = e^{-i \theta \sum_i (X_i+X_i^\dagger)}|0\rangle^{\otimes N}$, we study the evolution of mana on a subsystem of $N_A$ qutrits under a global brickwork circuit composed of random two-qutrit Clifford gates $u\in\mathcal{C}_2(d)$. 
As in the previous cases, the global magic is conserved under free operations, while the subsystem can dissipate resources as the circuit redistributes them. 
Our numerical results for dilution parameter $\epsilon=0.5$, with $N=12$ qutrits and $N_A=2$, are shown in Fig.~\ref{fig1:QME}(d). Similar to the non-Gaussianity case, the dynamics of the mana exhibits no crossings, indicating that tilted polarized states do not display a QME.

In summary, coherence and imaginarity exhibit clear signatures of the QME, while non-Gaussianity and magic do not when initialized in tilted states. Remarkably, even in these latter cases a preheating stage accelerates dissipation, inducing a 
QPME, as discussed in the following section.

\section{Pontus-Mpemba Effects} 
We now fix a resourceful state $|\Psi(\theta)\rangle$ and study how a preheating stage modifies its dynamics. 
For a 
circuit depth $t\leq T$, we consider a state evolved first via Haar-random circuits acting independently on subsystems $A$ and $B$, i.e.
$|\tilde{\Psi}\rangle = \tilde{U}^A_t \otimes \tilde{U}^B_t|\Psi(\theta)\rangle$, where $\tilde{U}^{A/B}_t=\prod_{s=1}^t \tilde{u}_s^{A/B}$. Here, each layer is given by $\tilde{u}^J_s = \prod_{x\in \mathcal{B}^{J}_s} \tilde{u}_{s,x,x+1}$, acting separately on $J=A,B$ with $\mathcal{B}^{J}_s= \mathcal{B}_s\cup J$, and the two-qudit gates  $\tilde{u}_{s,x,x+1}$ are sampled independently from the Haar measure on $U(d^2)$. 
For convenience, we introduce two dilution parameters, $\epsilon_A$ and $\epsilon_B$, controlling the strength of the preheating in each subsystem. After this stage, the dynamics 
is generated by the circuit $U_t$ comprising the free gates~[cf. Fig.~\ref{fig:schematic}(b)].

\begin{figure}[t!]
\centering
\includegraphics[width=0.5\hsize]{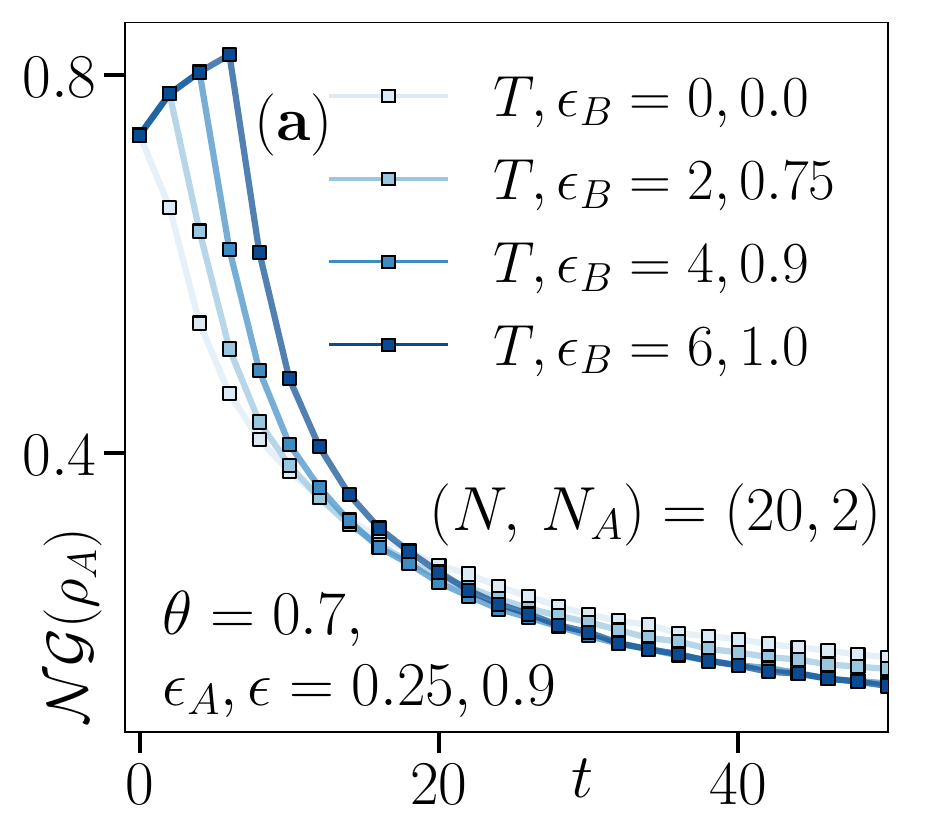}\hspace{-0.01\hsize}%
\includegraphics[width=0.5\hsize]{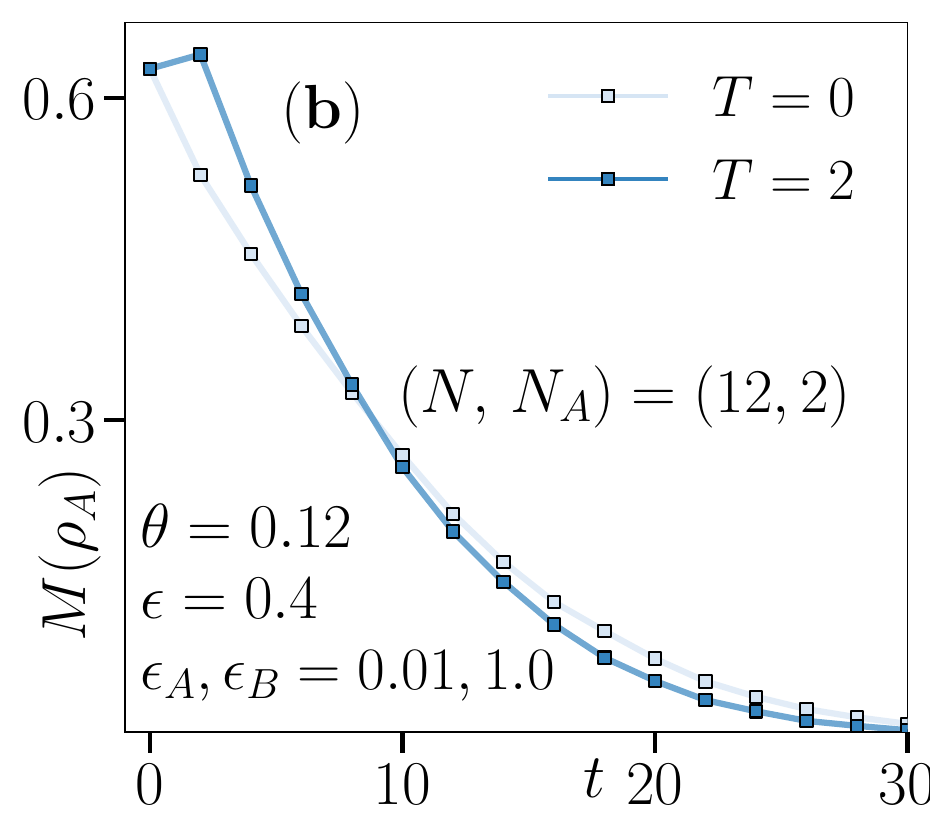}\\
\caption{The QPME for quantum resources in a 1D chain of $N=20$ qubits: non-Gaussianity (a); and for $N=12$ qutrits for quantum magic resources (b), for subsystem size $N_A=2$. 
The initial state is $\ket{\Psi(\theta)}$ and the dynamics is now driven by a two-step protocol: first, with the resourcefulness increasing circuit $\tilde{U}^{A}_{t}\otimes \tilde{U}_{t}^{B}$ of depth $T$ and then, with $U_{t}$, comprising resource-free gates. 
All the results are averaged over $4000$ circuit realizations. 
We observe a faster decay of the local resource content for increasing $T$ --- the signature of the QPME.
}
\label{fig2:PQME}
\end{figure}

We first consider the setup for non-Gaussian evolution, fixing the initial state as $\ket{\Psi(\theta)}$ with $\theta=0.7$. 
The system is evolved under preheating for $t\leq T$ and subsequently under free dynamics for $t>T$. 
Our results, shown in Fig.~\ref{fig2:PQME}(a) for $\epsilon_A=0.25$, $\epsilon=0.9$, varying $0\leq T\leq 6$, $0\leq \epsilon_B\leq 1$, and system size $N=20$ qubits ($d=2$) with $N_A=2$, reveal that the preheating initially increases the local resource content, as expected. 
Once the heating is turned off, the system evolves freely.
The prior preheating accelerates the decay of non-Gaussianity $\mathcal{NG}$. 
In particular, clear crossings emerge as a function of the preheating depth and strength. These results highlight the phenomenology of the Pontus-Mpemba effect, previously studied in Lindbladian dynamics and~\cite{nava2025pontus} here observed in quantum complexity metrics.

Following a similar prescription, we now turn our focus on the magic resources. 
Fixing the initial state $|\Psi(\theta) \rangle = e^{-i \theta (X+X^\dagger)}|0\rangle^{\otimes N}$ for a system of $N=12$ qutrits and $N_A=2$, we consider the effect of preheating for $T=0,2$ $\epsilon_A=0.01$, $\epsilon_B=1.0$ and $\epsilon=0.4$, cf. Fig.~\ref{fig2:PQME}(b).
Again, the two-step resourceful-free protocol drives $\rho_A$ to relax faster than in purely free evolution, a hallmark of the QPME.

In the End Matter, we demonstrate that similar results apply even in the setups where a QME was already present, namely for the coherence and imaginarity. 
Taken together, our findings indicate that the 
QPME is a 
ubiquitous feature of quantum resources: when the QME is absent, preheating can induce it, and when it is present, preheating maintain it.

\section{Conclusion} 
In this Letter, we investigated the quantum Mpemba effect through quantum resources that capture aspects of many-body quantum complexity.
Specifically, we examined coherence, non-Gaussianity, magic, and imaginarity, thereby extending the Mpemba paradigm beyond its traditional thermal and symmetry-based settings. 
We demonstrated that, for suitable initial states, Mpemba-like behavior arises when the reduced dynamics is monitored via appropriate resource monotones. 
Furthermore, even for classes of states where no Mpemba effect is normally observed, a preheating stage induces the recently introduced quantum Pontus-Mpemba effect.

Our results open several avenues for future investigation. First, the phenomenology observed here, supported by extensive numerics, calls for an analytical framework that can provide a physical understanding, in the spirit of recent progress on asymmetry. This is a particularly challenging task, as the systems considered here offer no immediate connection to physical degrees of freedom, such as quasiparticle descriptions~\cite{rylands2024microscopic}, and we leave it for future work. Another promising direction is to analyze the microscopic origin of these effects within recently discussed Markovian evolutions~\cite{summer2025resourcetheoreticalunificationmpemba,Bao2025mpembareset}. A preliminary study in this direction is presented in the End Matter for the theory of quantum coherence, but a systematic analysis lies beyond the scope of the present work.

Since the quantum resources discussed here play a fundamental role in synthetic quantum matter, it is natural to ask how monitored dynamics~\cite{fisher2023random,Potter_2022,li2025measurementinduced} affect the emergence of Mpemba physics, cf. recent work on this direction on asymmetry~\cite{DiGiulio2025measurement,travaglino2025quenchdynamicsentanglemententropy}.
More broadly, our results show that more resourceful systems can dissipate their local resources faster. It is an intriguing question whether this quantum and Pontus Mpemba effects in complexity can be harnessed for practical purposes in current and near-term quantum devices~\cite{nava2025speedingpontusmpembaeffectsdynamical}. We leave these and related questions as an outlook for future research.

\begin{acknowledgments}
\section{Acknowledgments} 
We thank L. Leone for discussions at the early stage of the project. We thank E. Tirrito, P. Calabrese, A. De Luca, J. Goold, S. Murciano, A. Nava, R. Egger for discussions. 
SA acknowledges support from Alexander von Humboldt foundation as a Humboldt postdoctoral fellow. XT acknowledges support from DFG under Germany's Excellence Strategy - Cluster of Excellence Matter and Light for Quantum Computing (ML4Q) EXC 2004/1 – 390534769, and DFG Collaborative Research Center (CRC) 183 Project No. 277101999 - project B01.
AS acknowledges the financial support provided by Microsoft Ireland.
PS acknowledges fellowship within the “Generación D” initiative, Red.es, Ministerio para la Transformación Digital y de la Función Pública, for talent attraction (C005/24-ED CV1), funded by the European Union NextGenerationEU funds, through PRTR.

\paragraph{Code and Data Availability.} The code and the data for our simulations will be publicly shared at publication. 

\paragraph{Note Added:} During the completion of this work, a parallel contribution appeared on the Pontus--Mpemba effect in the context of asymmetry theory for Hamiltonian systems~\cite{yu2025quantumpontusmpembaeffectsreal}. While the scope of our study is markedly different, the results agree in the overlapping regime.  
\end{acknowledgments}

\bibliography{ref.bib} 

\newpage 
\begin{center}
{\bf End Matter}
\end{center}

\begin{figure*}[t!]
\centering
\includegraphics[width=0.25\textwidth]{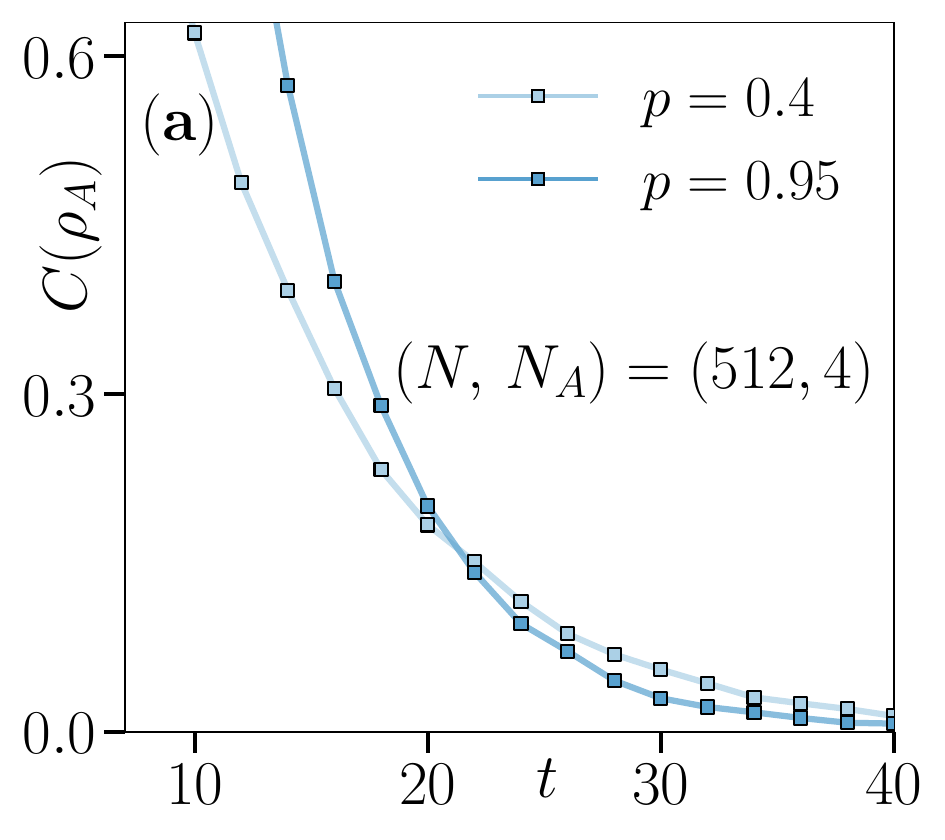}\hspace{-0.005\hsize}%
\includegraphics[width=0.25\textwidth]{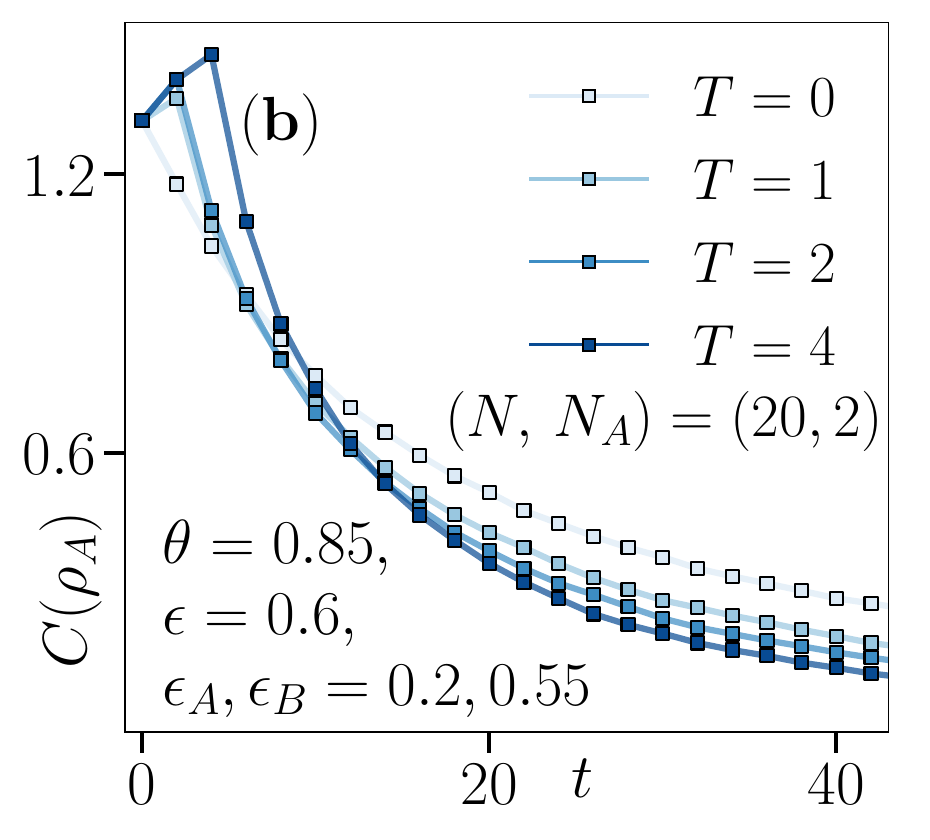}\hspace{-0.01\hsize}%
\includegraphics[width=0.25\textwidth]{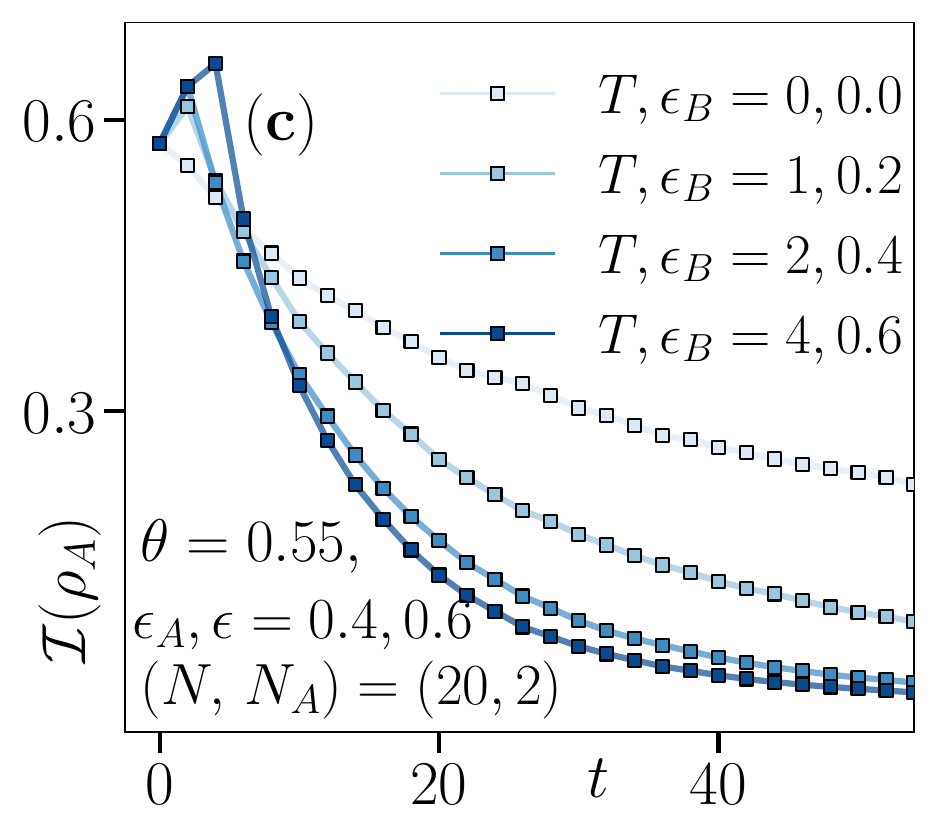}\hspace{-0.012\hsize}
\includegraphics[width=0.25\textwidth]{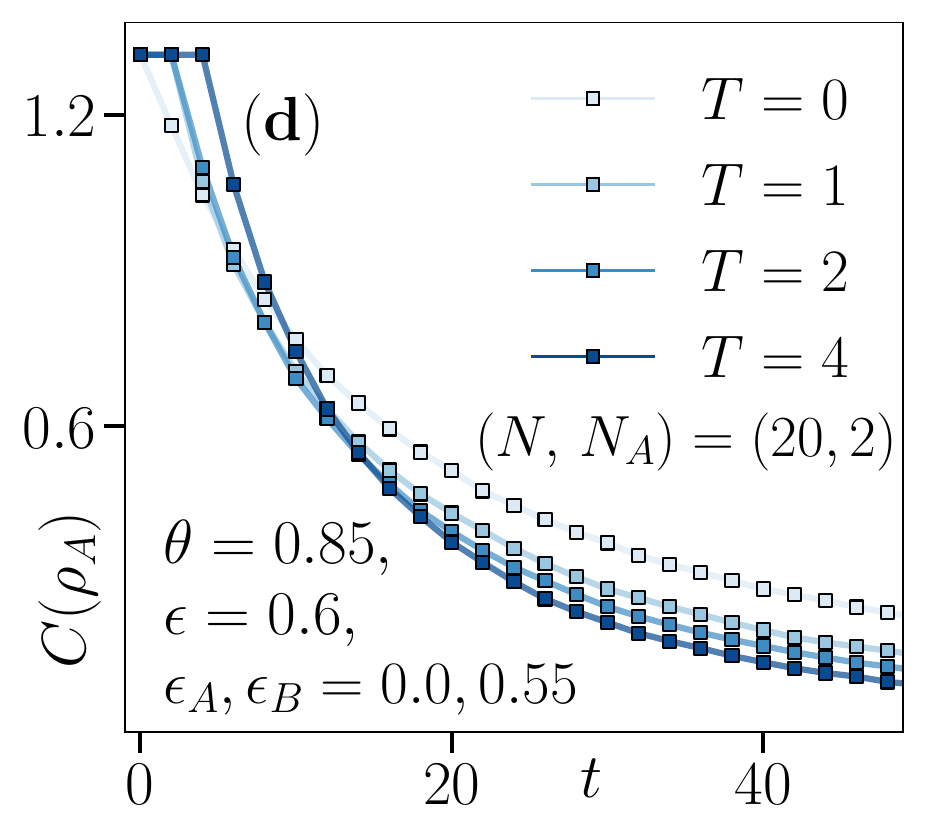}
\caption{ The QME for quantum resources in a 1D chain of $N=512$ qubits using the Clifford simulation and the QPME with exact vector simulation for $N=20$ for two resources: coherence (b) and imaginarity (c), and coherence again, however, with a slightly setup from the earlier one (d). In the first case, we monitor the resource content of a subsystem of size $N_A=4$ (a)  and $N_A=2$ for the last three cases (b-d), where the results are averaged over $4000$ disorder realizations. In Fig.~(a), the system is initialized with qubits in $\ket{+}$ (probability $p$) or $\ket{0}$ (probability $1-p$), and evolves under free operations, demonstrating the robustness of QME for large sizes. In Figs.~(b–c), starting from a common initial state $\ket{\Psi(\theta)}$, the system is evolving for $T$ layers with the circuit $\tilde{U}_{t}^{A}\otimes \tilde{U}_{t}^{B}$ including resourceful gates, then under free operations, yielding the characteristic QPME crossings for coherence and imaginarity. Finally, Fig.~(d) shows that preheating restricted to subsystem $B$ already suffices to capture the QPME crossing.}
\label{fig3:QME_clifford}
\end{figure*}

\section{Quantum Mpemba Effect in Coherence Resources via Clifford circuits}
We corroborate the main results presented in the Main Text studying the dynamics of initial resourceful state under coherence-preserving Clifford circuits. 
To that end, 
we consider an initial product state
preparing by applying a Hadamard gate $H=(X+Z)/\sqrt{2}$ with probability $p$ over each possible site. On average, this state has $k=p N$ states initialized in $|+\rangle=(|0\rangle + |1\rangle)/\sqrt{2}$, and the remaining $N-k$ qubits are in the state $|0\rangle$. 
Higher $p$ correspond to more resourceful initial states. 
We then process these initial states via a random Clifford circuits where each two qubit operations preserve coherences (768 gates out of the total $|\mathcal{C}_2(2)|=11520$ Clifford gates on two qubits). 
Crucially, the relative entropy of coherence $C(\rho_A)$ for the subsystem $\rho_A$ can be computed efficiently with the techniques for stabilizer states.

On the one hand, $S(\rho_A)$ is efficiently computable via the methodologies presented in~\cite{hamma2005bipartite}. 
The reduced density matrix of $A$ can be written as
\begin{equation}
\rho_A=\frac{1}{2^{N_A}}\sum_{\substack{g\in\mathcal{S}\ g|_B=I}} g|_{A},
\label{eq:red_den_matrix_stab}
\end{equation}
where the sum runs over stabilizers acting trivially on $B$. 
Denoting the domain of this reduced sum by $\mathcal{G}_A$,  Eq.~\eqref{eq:red_den_matrix_stab} can be rewritten as
\begin{equation}
\rho_{A}=\frac{|\mathcal{G}_{A}|}{2^{N_A}}\sum_{g\in \mathcal{G}_{A}} \frac{g}{|\mathcal{G}_{A}|},
\end{equation} 
where $\sum_{g\in \mathcal{G}_{A}} g/|\mathcal{G}_{A}|$ is the projector onto the subspace stabilized by $\mathcal{G}_{A}$. Hence, the entropy becomes $S(\rho_{A})=N_{A}-\log_2|\mathcal{G}_{A}|$. 
Given the $N$ generators $g_\mu=e^{i\phi_\mu} \prod_{i=1}^N X_i^{m_i^\mu} Z_i^{n_i^\mu}$ of the global state $|\Psi\rangle$, we consider the $N\times (2N+1)$ tableau matrix fixed by the elements $\phi^\mu$, $m_i^\mu$, and $n_i^\mu$. We consider the submatrix $\mathcal{M}_A=(m_i^\mu,n_i^\mu)_{\mu=1,\dots,N}^{i\in A}$ of size $N\times 2N_A$. Then one can prove the simplified $S(\rho_A) = \mathrm{rank}_{\mathbb{Z}_2}(\mathcal{M}_A)-N_A$, cf.~\cite{nahum2017quantum,sierant2023entanglement}. 
A similar method applies to the computation of $S(\rho_A^\mathrm{diag})$, where the diagonal part over the computational basis states $|z\rangle$ is
\begin{equation} \rho_{A}^{diag}=\sum_{z}\langle z|\rho_{A}|z\rangle \ket{z}\bra{z}\;.
\label{eq:rho_diag}
\end{equation}
This diagonal matrix is obtained further restricting the choices of operators to retain only stabilizers generated by $I$ and $Z$ operators on $A$, cf. Eq.~\eqref{eq:red_den_matrix_stab}, namely
\begin{equation}
    \rho_A^\mathrm{diag}=\frac{1}{2^{N_A}}\sum_{\substack{g\in\mathcal{S}\; g|_B=I\\ g|_A={I,Z}}} g|_{A}= \frac{|\mathcal{G}^\mathrm{diag}_{A}|}{2^{N_A}}\sum_{g\in \mathcal{G}_{A}} \frac{g}{|\mathcal{G}_{A}^\mathrm{diag}|}\,.
\label{eq:red_den_matrix_stab_diag}
\end{equation}
The restricted group $\mathcal{G}_A^\mathrm{diag}$ is encoded in the kernel of the sub-matrix $\mathcal{M}_A^\mathrm{diag}=(m_i^\mu,n_j^\mu)_{i=1,\dots,N\;, j\in B}^{\mu=1,\dots,N}$. 
Therefore, using standard linear algebra, we have
\begin{equation}
    S(\rho_A^\mathrm{diag})= N_A-\mathrm{dim}\;\mathrm{ker}(\mathcal{M}_A^\mathrm{diag})=\mathrm{rank}_{\mathrm{Z}_2}(\mathcal{M}_A^\mathrm{diag})-N_B. 
\end{equation} 
With all ingredients in hand, we afterward inspect the coherence content of the initial state under reduced dynamics over time with Clifford simulation as shown in Fig. \ref{fig3:QME_clifford} (a) for $N=512$ and $N_A=4$. This showcases 
another occurrence of the QME, with features identical to the result reported for much smaller systems in the Main Text.

\section{Additional Results on Pontus-Mpemba Effects}
Similarly to the resource theories of non-Gaussianity and magic resources, we show that 
QPME arises also in the coherence and imaginarity case.
The setup is the same considered in the Main Text for non-Gaussianity, for both these resources. 
First, let us consider the coherent case, and fix the tilted ferromagnetic state with $\theta=0.85$ as the initial state. 
We evolve the system with a preheating for $0\le T\le 4$ with $\epsilon_A=0.2$ and $\epsilon_B=0.55$ and then letting the system evolve with a free evolution with $\epsilon=0.6$. 
Our results, shown in Fig.~\ref{fig3:QME_clifford}(b) for system size $N=20$ qubits with $N_A=2$, reveal that preheating initially increases the local resource content, as expected. 
In Fig.~\ref{fig3:QME_clifford}(c) we present similar results for the imaginarity case. 
Concretely, we fix the initial state by $\theta=0.55$ and the evolution with $\epsilon_A=0.55$ and $\epsilon=0.6$, while $0\le T\le 4$ with $0\le \epsilon_B\le 0.6$. 
Again, once the heating is turned off, the system evolves freely, yet the prior preheating accelerates the decay of the relative entropy of imaginarity, with faster relaxation for stronger or deeper heating. 
These results corroborate the overall picture put forward in the Main Text.
Finally, we perform another analysis with a setup almost identical to Fig.~\ref{fig3:QME_clifford}(b), but this time restricting the preheating stage only to subsystem \(B\), which is not subjected to the monitoring for the reduced resource content. This case, shown in Figure~\ref{fig3:QME_clifford}(d), demonstrates that such a preheating protocol, applied prior to the resource-free gates, is sufficient to produce the characteristic QPME crossing, thereby extending beyond the original PME setup.

\begin{figure}[t!]
\includegraphics[width=0.5\hsize,height=3.7cm]{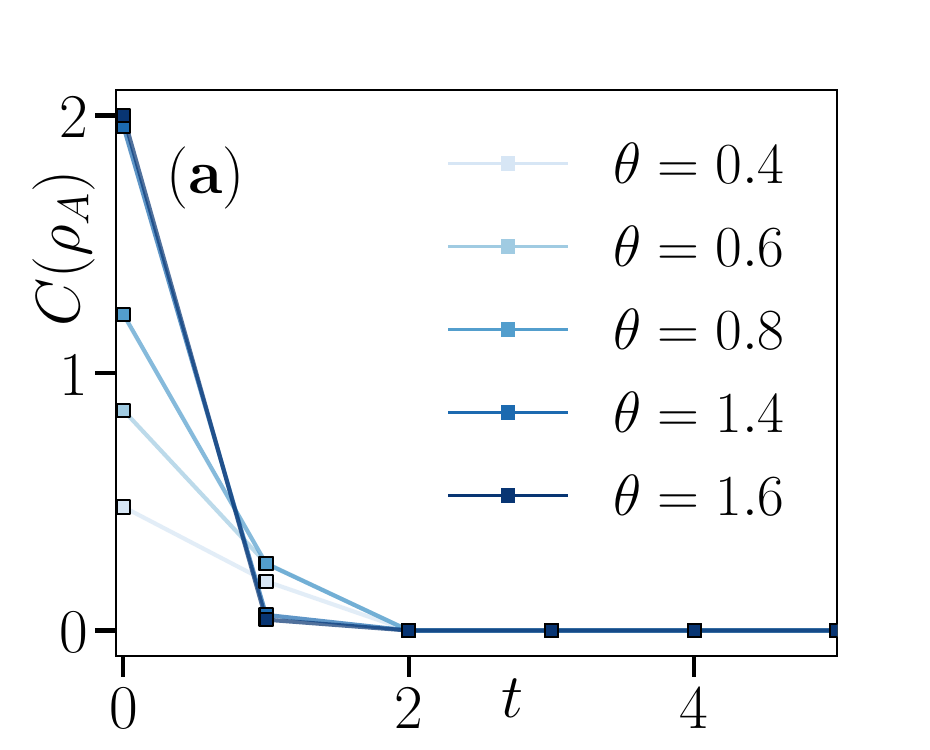}\hspace{-0.01\hsize}%
\includegraphics[width=0.5\hsize,height=3.7cm]{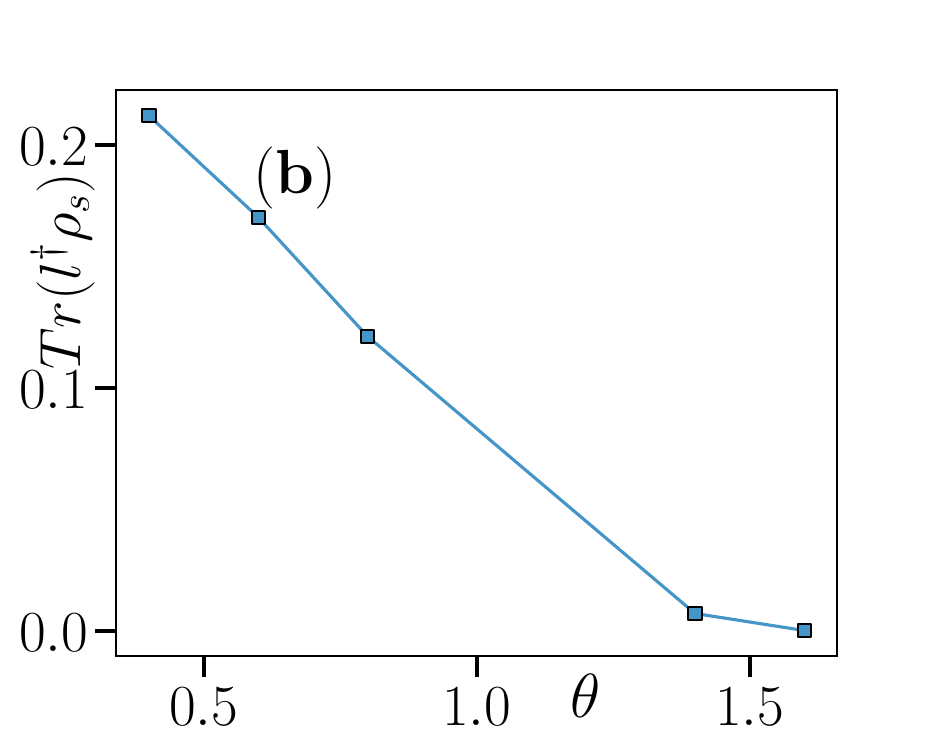}

\caption{Markov dynamics in case of Coherence-preserving Floquet evolution for $N=6$ and $N_A=2$ for a fixed gate realization in OBCs. Left panel: The time evolution of the resource monotone for the reduced dynamics of the subsystem showcasing the Mpemba effect. (b): The overlap of the slowest-moving mode of the superoperator with the initial state decreases with increasing resources, i.e., $\theta$-values. }
\label{fig:QME_markov}
\end{figure}

\section{Markovian Open Evolution}
As a preliminary step toward a systematic study of quantum complexity in open systems, and a search for a microscopic origin of Mpemba phenomenology, we consider a Markovian setup~\cite{Breuer2002openquantum,Gardiner2004openquantun}. 
Here a chain of $N$ qubits is partitioned in a system of interest of size $N_A$ and an environment on the remaining sites. 
The system is initialized in the state
\begin{equation}
\rho_s= (\ket{\theta}\langle \theta|)^{\otimes N_A} \otimes \frac{\mathbb{1}_B}{2^{N-N_A}},
\end{equation}
where $\ket{\theta} = e^{-i Y \theta/2} \ket{0}$ and the environment is in the 
maximally mixed state $\pi_e \equiv \mathbb{1}_B/2^{N-N_A}$ 
(which is a free state in the coherence resource theory). 

The full system evolves under a brickwall Floquet circuit composed of fixed two-qubit coherence-preserving gates $U_{x,y}$ in OBCs as
\begin{equation}
U = \prod_{x=1}^{N/2-1} U_{2x,2x+1} \prod_{x=1}^{N/2} U_{2x-1,2x}.
\end{equation}
The identical gate is chosen as $U_{x,x+1}=u$ with a fixed value 
\begin{equation}
\begin{split}
    u &= (0.79517062 - 0.60638575 i)\,|00\rangle\langle 11|\\&
+ (0.6283294 - 0.7779474 i)\,|01\rangle\langle 01|\\
&+ (0.99984225 + 0.01776183 i)\,|10\rangle\langle 10|\\&
+ (-0.96625112 - 0.25760196 i)\,|11\rangle\langle 00|.
\end{split}
\end{equation}

The evolution of the system can be described by the quantum channel as
\begin{equation}
\mathcal{E}[\rho_s] = \mathrm{Tr}_e \left[ U (\rho_s \otimes \pi_e) U^\dagger \right],
\label{Eq:quantum_channel}
\end{equation}
where 
$U$ comprises free unitaries, in analogy to the unitary dynamics setup considered in the Main Text.
Each application of the channel $\mathcal{E}$ is associated with tracing out and resetting the environment 
to the maximally mixed state $\pi_e$, rendering the dynamics memoryless (Markovian)~\cite{Breuer2002openquantum,Gardiner2004openquantun,summer2025resourcetheoreticalunificationmpemba}. 
The effective dynamics can further be recast as the quantum master equation
\begin{equation}
\rho_s(t) = e^{\mathcal{L}t} [\rho_s],
\end{equation}
where $\mathcal{L}$ is the (generally non-Hermitian) generator. Assuming $\mathcal{L}$ is diagonalizable, we denote its right and left eigenvectors and eigenvalues by
\begin{equation}
\mathcal{L} \ket{r_k} = \lambda_k \ket{r_k}, \qquad
\bra{l^{\dagger}_k} \mathcal{L} = \lambda_k \bra{l^{\dagger}_k}.
\end{equation}
The time evolution of any initial state can now be written as
\begin{equation}
e^{\mathcal{L}t}[\rho] = \rho_{ss} + \sum_{k=2}^{d^2} e^{t \lambda_k} \, \mathrm{Tr}(l_k^\dagger \rho) \, \ket{r_k},
\end{equation}
where $\rho_{ss}$ denotes the steady state with $\lambda_1 = 0$, $d=2^{N_s}$ in our case, and $\mathrm{Re}(\lambda_k) \le 0$ for all $k$. The slowest decaying mode is associated with $\lambda_2$ (in the sorted eigenvalue spectrum of the generator based on the real part of the eigenvalue) and controls the asymptotic decay at long times. For a specific choice of $\theta$-values, we find that the QME can be explained by the spectral decomposition of the eigenvalues of the superoperator. For this choice of initial states, the more resourceful state (larger $\theta$) is seen to have a smaller overlap with the slowest mode $\ket{r_2}$ than a lower-resource state, thus facilitating us to observe the QME in this setup, as shown in Fig. \ref{fig:QME_markov} (a-b) for $N=6$ and $N_A=2$.

\end{document}